\documentclass[overcite,prl,aps]{revtex}
\usepackage{graphics}

\begin{document}

\twocolumn[\hsize\textwidth\columnwidth\hsize\csname@twocolumnfalse\endcsname

\title{A Brief Comment on ``The Pyrotechnic Universe" }

\author{Justin Khoury$^1$, Burt A. Ovrut$^2$, Paul J. Steinhardt$^1$
and Neil Turok $^{3}$}

\address{
$^1$ Joseph Henry Laboratories,
Princeton University,
Princeton, NJ 08544, USA \\
$^2$ Department of Physics, University of Pennsylvania,
Philadelphia, PA 19104-6396, USA\\
$^3$ DAMTP, CMS, Wilberforce Road, Cambridge, CB3 0WA, UK}

\maketitle

\begin{abstract}
We respond to the  criticisms
by Kallosh, Kofman and Linde concerning our proposal of
the ekpyrotic universe scenario.  We point out a number of errors
in their considerations and argue that, at this stage,
the ekpyrotic model is a possible
alternative to inflationary cosmology as a description of the very
early universe.
\end{abstract}
\pacs{PACS number(s):  98.62.Py, 98.80.Es, 98.80.-k }
]

We have recently proposed an alternative to inflationary
cosmology,\cite{Gut,Lin}  entitled the ``ekpyrotic universe,"\cite{ek1}
designed to resolve the
horizon, flatness, and monopole puzzles of standard hot big bang
cosmology and to generate a nearly scale-invariant spectrum of adiabatic
energy density perturbations needed to seed structure formation.
The basic notion, motivated by string theory and M-theory,\cite{witten1,lukas1}
is that
the hot big bang universe is produced by the collision in an
extra-dimensional space-time between a  three-dimensional
brane in the bulk space with another brane or
a bounding orbifold plane.  The collision heats the universe
to a high but finite temperature, from which point the hot big
bang phase begins.
Whereas the inflationary scenario relies on an extended period of
exponential hyperexpansion prior to the hot big bang phase,
the ekpyrotic model relies
on extremely slow evolution over an exceedingly long time.

In a recent preprint,\cite{pyr} Kallosh, Linde, and Kofman (KKL)
raised a number of criticisms of the ekpyrotic universe. 
In this comment we briefly explain why we respectfully disagree
with each of their seminal
points (italics, below).
Where a full  response requires a
technical calculation, the details are given in
the separate publications to which we refer.

We first consider
the criticisms of the superstring and M-theoretic
underpinnings. We then discuss the description of
density perturbations in our scenario and finally,
the criticisms of the initial conditions.

\noindent
{\it  KKL: One of the central points of the ekpyrotic cosmological scenario
is that we live on a negative tension brane.  However, the tension
of the visible brane in Ho\v{r}ava-Witten
theory, as well as in relevant cases
of non-standard embedding, is positive. Hence, there is a problem
in the assignment of signs to the brane tensions.}

This criticism is factually incorrect.
The claim is  based on experience with  the standard embedding and
other specific examples that have appeared
in the literature previously,
rather than on mathematical  analysis.  The fact is
that there never was a  cohomology
requirement that the visible brane have
positive tension to obtain realistic models.
We have constructed numerous examples\cite{bao} of heterotic M-theory
models in which the visible brane has negative tension and is endowed
with a much smaller gauge group ({\it e.g.}, SU(5)) than that of the 
positive tension hidden brane ({\it e.g.}, $E_7$).

It is the standard embedding that motivated the minor variant 
suggested by  Kallosh, {\it et al}, dubbed by them
the ``pyrotechnic" universe.  Ironically, this particular choice
is  disallowed
since bulk branes are mathematically
forbidden, and so there could
no brane collision to ignite the hot big bang phase.

We should also emphasize that,
contrary to KKL's assertion, we never claimed that the
assignment of negative tension to the visible brane is a
``central point" of the ekpyrotic concept.
We did not intend our proposal
to be interpreted so narrowly.
It is  a feature of our specific example, and
Ref.~\ref{bao} demonstrates that this example is
possible. For this example, we explained
the role that negative tension
plays in the energetics, fluctuations, and ultimate
expansion of the universe.  But, we also stated that
the general principles could be adapted to
numerous set-ups.\cite{ek1}
Indeed, in Refs.~\ref{ek1} and~\ref{ek2}, we derive the perturbation
spectrum in a 4d static background in a limit which
is totally insensitive to the assignment of brane tensions.
The Kallosh {\it et al.} example, which is based on the
same physical principles and mathematical equations,
but  entails flipping a few signs in the action,
has no substantive differences to our original scenario
(provided the embedding problem is fixed).

Kallosh {\it et al.}  not only criticize the choice of sign for the
tensions, but also the magnitudes:

\noindent
{\it KKL:
In examples considered in the literature, the contribution
to the cohomology constraint
from the bulk brane is
of the same order as the one from the boundaries,
whereas the ratio is extremely small $(4 \times 10^{-5})$
in the ekpyrotic model.}
More precisely,
the claim is that the ratio of the bulk tension $\beta$
to the  boundary tension $\alpha$ in our
example is incompatible for realistic models
with what is required by the cohomology constraint.

This criticism again rests
on  examples of Ho\v{r}ava-Witten  models
in the literature rather than on mathematical analysis.
In Ref.~\ref{bao}, it is shown that there is a wide
range of freedom for $\beta/\alpha$ and that the value
in our example is mathematically consistent with
the cohomology constraint.

Furthermore, we did not intend
our proposal to be
interpreted so narrowly. Let us not confuse an example
with general principles.
We never claimed that a small ratio is
required. Indeed, viable and observationally
consistent examples with ratios greater  than  1/10
are discussed in Ref.~3.

Other criticisms about the string-theoretic underpinnings
include:
\begin{itemize}
\item {\it questioning the presence of the 4-form in the
heterotic M-theory action}: 
The 4-form formulation of the action is equivalent to
the action presented in Ref.~(\ref{lukas1}).  This is easily
seen by eliminating the 4-form using its equation of motion. This
formulation is particularly useful in heterotic vacua with bulk branes, such as
in ekpyrotic cosmology.
\item {\it questioning the origin of the bulk 3-brane}:
The 3-brane is simply an M5 brane
 wrapped on a holomorphic curve of the Calabi-Yau threefold.
\item {\it stabilization of moduli}:  Granted, this is
a deep, unsolved problem of string theory.  We are
presuming its solution does not interfere with the
ekpyrotic scenario. The same presumption must be made
in inflationary theory.
Indeed, in this case,
there are well-documented difficulties that
arise with inflation if moduli are not stabilized 
prior to inflation (which may prevent inflation occurring
 \cite{moduli}), 
difficulties which are not applicable
to the ekpyrotic model.
\end{itemize}

In addition, Kallosh {\it et al.} criticize
the sign of the potential,  its parameters and
the notion that the potential is zero after collision.
Granted, assuming the potential to be zero after
collision is fine-tuning.
However, let us recognize it for what it is -- the
well-known cosmological constant problem.
Precisely the same fine-tuning of the true vacuum 
energy
is
required in inflationary cosmology.

As for criticisms about tuning of parameters,
this skepticism is based largely on a search of
the current literature, all of which was written
prior to the appearance of the ekpyrotic proposal,
and on the one example in our paper.
There is no serious physics argument, or mathematical
analysis,  or analytical discussion
of what range of parameters is required or what
emerges naturally from string theory. Indeed, our
choice of parameters and potential appears to 
us to
be well within the
bounds one might expect from string theory.  However,
to make their point,  Kallosh {\it et al.}
reparameterize the action in variables that are not
natural to string theory and then argue that the
potential is finely-tuned. 
This seems a strange and unnecessary procedure.

More generally, any conclusions about
parameters are premature since
there has been no serious analysis of what range
is  required other than the point made
in our paper Ref.~\ref{ek1} that there appears to be
a great deal of flexibility.
Perhaps some historical perspective is useful.
A similar
search of the literature prior to first computations
of density perturbations in 1982 would reveal no examples
of slow-roll potentials with dimensionless couplings
of  order
$10^{-14}$.  History shows that inflationary cosmology
stimulated the search for these models, and now many
examples are found in the literature.   We would
suggest that some patient, serious analysis is required
before the ekpyrotic scenario can be judged
against inflation in this respect.

\noindent
{\it KKL: The mechanism for the generation
of density perturbations in
this scenario is a particular limiting
case of the mechanism of
tachyonic preheating.}

The basic notion of the ekpyrotic model is that the
universe begins in a quasi-static (non-expanding) state,
a concept that dates back to ancient philosophy.
In pursuing this idea, a basic challenge is to generate
fluctuations which are in accord with the impressive
and precise
measurements of the cosmic microwave background and
large-scale structure which lend strong support
for a nearly scale-invariant spectrum of linear, 
adiabatic density perturbations.
Our paper shows that such perturbations are generated 
in a quasi-static multi-brane universe
for certain
simple potentials 
including negative exponentials and inverse power
laws. 

Tachyonic preheating,\cite{feld} a concept  introduced
by Kofman, Linde and collaborators,
concerns phenomena not directly related to the
generation of large scale density perturbations, in a rapidly
expanding universe just following inflation.
It is legitimate within its own context. 
But the potentials needed for scale invariance do not 
appear anywhere in the tachyonic preheating paper,
and we think that
it obfuscates rather than clarifies the issue 
to conflate our fluctuation generation mechanism 
with tachyonic pre-heating. There 
 is no more relation  to
 tachyonic preheating than there  is to
 fluctuation generation in inflation. 
 In both cases, there
are some common elements but also major differences.

\noindent
{\it KKL: Inflation removes all previously existing
inhomogeneities and is
robust, whereas the ekpyrotic model  makes
the homogeneity problem
much worse.}

The critique refers to the fact that inflation is a mechanism
that can make an inhomogeneous universe more homogeneous, whereas
homogeneity in the ekpyrotic model is part of the initial 
condition.
Behind this critique lies a substantive point concerning the
different assumptions of
the inflationary and ekpyrotic scenarios.
However, a more objective  comparison is warranted.

Inflation assumes that the universe begins in a high energy 
state of no particular symmetry that is rapidly expanding from the 
start.  For such general initial conditions, it is essential 
that there be a dynamical attractor mechanism that makes the 
universe more homogeneous as expansion proceeds.  Superluminal
expansion provides that mechanism.
The ekpyrotic model, on the other hand, is built on the
principle that the initial state is quasi-static
with properties  dictated by symmetry, particularly supersymmetry.
So, by construction, the initial state is special, but special
in a way that is well-motivated physically.  
In this case, while
a dynamical attractor mechanism may be possible (see below),
it is not essential.
One can envisage the possibility that
the initial conditions are simply  the result of
some selection rule that dictates a state of maximal supersymmetry 
and low energy.

Within the context of superstring theory and M-theory, the natural
choice of initial state
is the BPS (Bogolmon'yi-Prasad-Sommerfeld) state.\cite{lukas1}
The BPS property is already
required from  particle physics in order to have a
low-energy, four-dimensional
effective action with ${\cal N}=1$ supersymmetry,
necessary for  a realistic
 phenomenology.  For our purposes,
 the BPS state is ideal because,
not only is it homogeneous,  as one might suppose,
but it is also flat.  That is, the BPS condition
links curvature and  homogeneity. 
It requires the two boundary branes to be parallel.

We do not rule out the possibility of a dynamical attractor 
mechanism that drives the universe towards the BPS state from 
some more general initial condition.
Such parallelism would  be a natural consequence 
of all the branes emerging from one parent brane \cite{nosing}.
Alternatively, one can imagine beginning with two
boundary branes only and no bulk-brane or
interbrane potential.
Perhaps curvature can  be dissipated by radiating excitations
tangential to the branes and  having them travel off to infinity.
The branes might settle into a BPS ground state  until, through
some rare process, a bulk brane is nucleated. Depending on the gauge
structure on the branes, the bulk brane may be drawn towards the
visible brane.  

Beginning in an empty,  quasi-static  state solves some problems, but
one should not underestimate the
remaining challenges:
how to generate a hot universe, and how to generate 
perturbations required for large-scale
structure. The remarkable feature of  the ekpyrotic picture is
that brane collision naturally
serves both roles. 

Inflationary cosmology, based upon a powerful 
attractor mechanism,
is very appealing philosophically, and we do not seek
to discredit it by proposing an alternative. 
But,  we would hesitate to characterize inflation as ``robust."
If robustness means that a Universe
such as the one we observe is {\it nearly inevitable}
in the context of an inflationary model, one must disagree. 
The presence of large initial inhomogeneities and black holes can 
always prevent inflation from occurring. 
There is no rigorous or even postulated measure
of the basin of attraction.  Claims that inflation works for
generic chaotic initial conditions are simply ill-defined.

There are other longstanding, unresolved issues.
A successful 
inflationary scenario requires setting couplings of the
inflaton to itself and to other fields at levels more than 12 orders
of magnitude below the natural expectation.\cite{turner}
Furthermore, inflation is far from being a rigid theory.
While flatness is certainly the simplest possibility,
at least one of the authors of \cite{pyr} 
has accepted that 
open universes \cite{bucher} are also possible \cite{open}.
Rigorous methods
for deciding the relative probability of  patches with different
physical properties ({\it e.g.}, open versus flat) 
are not known.\cite{guth2}
Finally, inflation has not yet been embedded in
a theory of quantum gravity in a consistent manner.
Inflationary quantum fluctuations emanate from
exponentially sub-Planckian scales, and the physics on such
scales is very incompletely understood.
All of these features suggest that it may be something
of an exaggeration to describe inflation as
``robust'' at this stage.

We do not raise these points to argue
that inflation is flawed or should be discarded. 
On the contrary, inflation is an outstandingly successful
and appealing idea. 
However,
if one is comparing the inflationary and ekpyrotic approaches
to the homogeneity and flatness problems, it is  sensible to maintain
a fair and open-minded attitude, and to present the features and
unsettled issues in both cases as objectively as possible.

\vspace*{.2in}

We see no reason to rush to judgment.
Inflationary theory is based on quantum field theory, a well-established
theoretical framework. The model took several years to develop,
and it has been carefully studied and
vetted for twenty years. 
It is still in many respects incomplete.
Our proposal is based on unproven ideas
in string theory and entails many
features which are new.
A sober examination of the ekpyrotic approach  will
prove useful
whichever way it turns out.
If the model is found to be theoretically inconsistent,
then our confidence in the inflationary solution will
be enhanced.
If the model proves to be a legitimate alternative, then
there is a new possibility   to consider and
nature's choice will be decided
observationally by measurements of the
gravitational wave background.\cite{ek1}


\begin{thebibliography}{9999}


 \bibitem{Gut}  A. H. Guth, {\it Phys. Rev. D}{\bf 23}, 347, (1981).
\bibitem{Lin} A. D. Linde, {\it Phys. Lett.} {\bf 108B}, 389 (1982);
  A. Albrecht and P. J. Steinhardt, {\it Phys. Rev. Lett.} {\bf 48},
  1220 (1982).
\bibitem{ek1} J. Khoury, B.A. Ovrut, P.J. Steinhardt and N. Turok,
hep-th/0103239.  \label{ek1}
\bibitem{witten1} P. Ho\v rava and E. Witten,
   Nucl. Phys. {\bf B460} (1996) 506; {\bf B475} (1996) 94.
\bibitem{lukas1} A. Lukas, B.A. Ovrut and D. Waldram, Nucl. Phys. {\bf B532}
(1998) 43;  Phys. Rev. D {\bf 57} (1998) 7529; A. Lukas, B.A. Ovrut,
K.S. Stelle and D. Waldram, Phys. Rev. D {\bf 59} 086001 (1999).  \label{lukas1}
\bibitem{pyr} R. Kallosh, L. Kofman, and A. Linde,
hep-th/0104073.
\bibitem{bao} R. Donagi, J. Khoury, B.A. Ovrut,
P.J. Steinhardt and N. Turok,
in preparation. \label{bao}
\bibitem{moduli} R. Brustein and P. J. Steinhardt, {\it Phys. Lett. B}
{\bf  302}, 196 (1993);
T. Banks, M. Berkooz, G. Moore, S. H. Shenker, P. J.
 Steinhardt, {\it Phys.Rev. D}{\bf 52}, 3548 (1995);
G. Huey, B.A. Ovrut, P.J. Steinhardt and D. Waldram,
{\it Phys.Lett. B}{\bf 476}(2000) 379-386, (2000).
\bibitem{feld} G. Felder, J. Garcia-Bellido, P.B. Greene, L. Kofman,
A. Linde, and I. Tkachev, hep-ph/0012142.
\bibitem{iflucs} 
J.M. Bardeen, P.J. Steinhardt and M.S. Turner, Phys. Rev.
D {\bf 28} (1983) 679;
A.H. Guth and S.-Y. Pi, Phys. Rev. Lett. {\bf 49} (1982) 1110;
 S. W. Hawking, {\it Phys. Lett. B}{\bf 115}, 295 (1982).
 V. Mukhanov and G. Chibisov, {\it Pis'ma Zh. Eksp. Teor. Fiz.} {\bf 33},
 549, (1981);
 A. A. Starobinskii, {\it Phys. Lett. B}{\bf 117}, 175 (1982).
\bibitem{ek2} J. Khoury, B.A. Ovrut, P.J. Steinhardt and N. Turok,
in preparation.
\label{ek2}
\bibitem{nosing} J. Khoury, B.A. Ovrut, N. Seiberg,
P.J. Steinhardt,  N. Turok, in preparation.
\bibitem{turner} P.J. Steinhardt and M.S. Turner,
{\it  Phys. Rev. D}29, 2162-71 (1984).
\bibitem{bucher} M.A. Bucher, A.F. Goldhaber and N. Turok,
{\it Phys. Rev. D}{\bf D52}, 3314 (1995).
\bibitem{open}
A. Linde, {\it Phys. Rev. D}{\bf 59},   023503 (1999);
J. Garcia-Bellido and A. Linde, {\it Phys. Rev. D}{\bf 55},  7480 (1997).
\bibitem{guth2} A. Guth,  astro-ph/0101507.

\end{thebibliography}
\end{document}